\documentclass[authoryear]{sigplanconf}

\usepackage{amsmath}
\usepackage{graphicx}
\usepackage{url}

\begin{document}

\setlength{\pdfpageheight}{\paperheight}
\setlength{\pdfpagewidth}{\paperwidth}

\conferenceinfo{PROMOTO '13}{October 26, 2013, Indianapolis, IN, USA} 
\copyrightyear{2013} 
\copyrightdata{}
\doi{}




\titlebanner{banner above paper title}        
\preprintfooter{short description of paper}   

\title{MIT App Inventor}
\subtitle{Enabling personal mobile computing}

\authorinfo{Shaileen Crawford Pokress}
           {MIT Media Lab}
           {shaileen@media.mit.edu}
\authorinfo{Jos\'e Juan Dominguez Veiga}
           {MIT Medial Lab}
           {josmasflores@gmail.com}

\maketitle

\begin{abstract}
MIT App Inventor\footnote{\scriptsize{http://appinventor.mit.edu}} is a drag-and-drop visual programming tool for designing and building fully functional mobile apps for Android. App Inventor promotes a new era of \emph{personal mobile computing} in which people are empowered to design, create, and use personally meaningful mobile technology solutions for their daily lives, in endlessly unique situations. App Inventor\textquoteright s intuitive programming metaphor and incremental development capabilities allow the developer to focus on the logic for programming an app rather than the syntax of the coding language, fostering digital literacy for all.  Since it was moved from Google to MIT, a number of improvements have been added, and research projects are underway.
\end{abstract}




\section{Introduction}

\begin{figure}
\begin{center}
\includegraphics[scale=0.25]{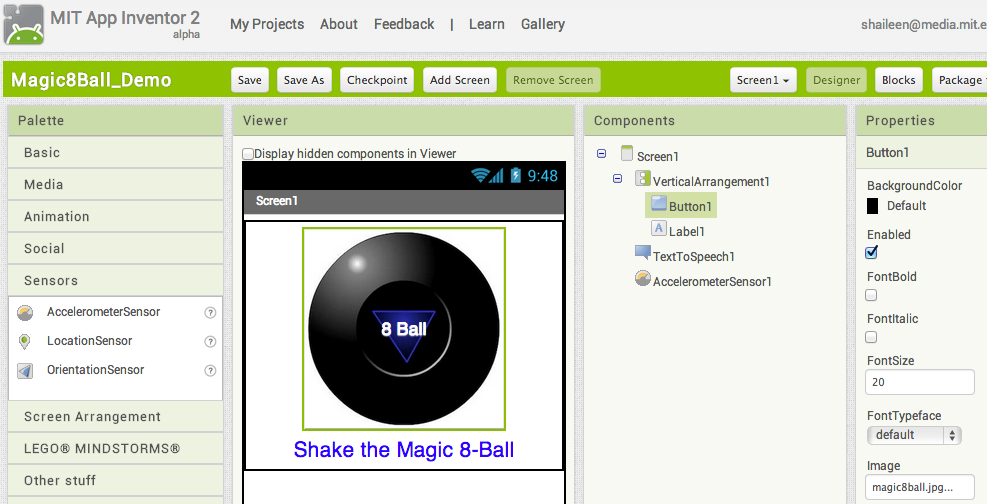}
\end{center}
\caption{The Designer provides an intuitive interface for choosing the app\textquoteright s components and their properties.}
\label{fig:designer}
\end{figure}

\begin{figure}
\begin{center}
\includegraphics[scale=0.35]{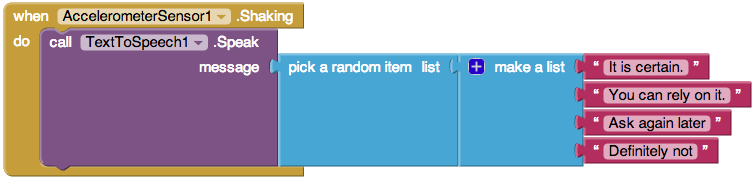}
\end{center}
\caption{The Blocks Editor gives graphical cues for programming the app\textquoteright s behavior. In this case, a yes/no prediction game.}
\label{fig:blocks}
\end{figure}

\paragraph{}
App Inventor\textquoteright s user interface is based on the idea of low-floor, high-ceiling development environments \citep{papert:mindstorms}, and consists of two parts: a {\bf{Designer}} (Figure \ref{fig:designer}) for selecting the components of the app, and a {\bf{Blocks Editor}} (Figure \ref{fig:blocks}) for setting the behavior of the app. App Inventor\textquoteright €™s building blocks are common user interface elements (buttons, labels, list pickers, images, etc.) coupled with the mobile device\textquoteright €™s features (texting, GPS, NFC, Bluetooth, etc.) Therefore, the primitive structures of the language enable the app developer to easily manipulate the functionalities of these touch-enabled, portable, sensing devices. 

\paragraph{}
By focusing on the device\textquoteright s functionality, App Inventor provides an intuitive programming metaphor. To build an app that sends and receives texts, a \emph{Texting} component can be used. To have an app respond to a user shaking the device, the \emph{Accelerometer} component can be used. Programming the app\textquoteright s behavior with blocks is just as intuitive: the block for handling shaking is called \textquotedblleft when Accelerometer.Shaking\textquotedblright . The block for detecting an incoming text is \textquotedblleft when Texting.MessageReceived\textquotedblright . This understandable, action-based, event-driven programming model reduces the frustration level from what is often experienced in traditional text-based programming environments.

\paragraph{}
Professor Hal Abelson of MIT developed App Inventor while on sabbatical at Google. He and a team of Google engineers launched the first version in 2009 as a project of Google Labs. Late in 2011, with a seed grant from Google, Abelson brought the project to MIT where it is now housed in the new Center for Mobile Learning (CML) at the Media Lab. CML  is co-directed by Abelson together with Mitchel Resnick (who leads the MIT Media Lab\textquoteright s Lifelong Kindergarten group, the home of Scratch) and Eric Klopfer (who leads MIT\textquoteright s Scheller Teacher Education Program, home of StarLogo TNG). 

\paragraph{}
In just 18 months since being re-launched by MIT, App Inventor has attracted over two million registered users. The server sees 40,000+  weekly active users, with a breakdown of users from all over the world. (Top 8 countries: US 18\%; UK 17\%; Taiwan 7\%; Italy 6\%; Spain 5\%; Brazil 5\%; Germany 4\%; Japan 3\%)

\section{Building Apps with App Inventor}

\paragraph{}
The App Inventor team has intentionally given the tool\, \textquotedblleft wide walls \textquotedblright\, \citep{resnick:walls} so that the audience and their creations can emerge organically. Unlike authoring tools that are specifically for making games or simulations, App Inventor provides building blocks that can be arranged for any purpose or goal. The tool itself is an access point to most of the features in a mobile device. In short, App Inventor opens up the\, €\textquotedblleft black box\textquotedblright\, that some perceive a handheld computer to be.

\subsection{Real-time Testing and Incremental Development}
\paragraph{}
One of the characteristics of working with App Inventor is that developers can see their creations while they are building them. This allows users to develop their apps \emph{incrementally} and encourages them to  test as they build. Whenever the user drags a new component to the designer, or creates new functionality through blocks, those new artifacts are automatically and readily available in the connected device or emulator. Users can select a particular block and see the results of the operation immediately by using the \emph{Do It} command from a contextual menu. This has very positive implications related not only to the benefits of the rapid feedback they receive, but also as a way of testing and debugging their creations while they are being made. Once the app is completed, it can be directly downloaded to the connected device, or exported in \emph{apk} format for distribution or upload to Google Play.

\section{The App Inventor User Community}
\paragraph{}
MIT App Inventor is used by students, teachers, developers, hobbyists, and entrepreneurs to develop apps for collaboration, productivity, personal use, recreation, learning, social good, and community activism. 
\paragraph{}
\emph{Educators and Learners} are a primary audience as App Inventor is first and foremost an academic project. Efforts to support educators in using App Inventor have resulted in a host of remarkable projects engaging students in app building, both within and outside of school walls. App Inventor\textquoteright s library of blocks include the most basic computational constructs such as conditionals, procedures, variables, and data structures. College and high-school faculty have successfully used App Inventor in their courses for over four years \citep{gray:csprinciples,wolber:realworld,ericson:camps,morelli:ct}, and there is a new Computer Science Principles course based on App Inventor.\footnote{\scriptsize{http://mobile-csp.org}} Indeed, App Inventor is an ideal environment for thinking and engaging computationally as evidenced by the collection of stories showcasing its successful use.\footnote{\scriptsize{http://appinventor.mit.edu/explore/stories.html}}
\paragraph{}
\emph{Hobbyists} enjoy creating and playing with technology, and App Inventor provides them a way to do that with mobile devices. There is a large community of people building apps for personal or community purposes. Several online user groups have sprung up around these interests. 
\paragraph{}
\emph{Entrepreneurs} are those who see App Inventor as a way to easily enter into the Android app market without having to learn to program in Java. This group creates business, entertainment, and utility apps that are posted on the Android app market, some for free and some for sale.
\paragraph{}
\emph{Contributors and Developers} are people who have taken it upon themselves to contribute code to the open source project or to build supporting materials for other users. They do this out of enthusiasm for the project and because they see the power of democratizing access to technological creation.

\section {Technical Considerations}
\paragraph{}
App Inventor apps only run on Android. Focusing on a single operating system provides more functionality to the end user than would be possible with a cross-platform solution. For example, App Inventor is tied into the Android OS in such a way that it can send and receive information from many of the components of the phone (GPS, NFC, texting, camera, accelerometer, etc.) In a cross-platform tool this would not be possible on all devices because of how different operating systems call these functionalities. Additionally, it is important that students be able to share their apps in a publicly accessible place without having to buy a developer license. The Google Play Store allows this, and in fact many App Inventor apps are now available in the Play Store.

\paragraph{}
App Inventor is an Open Source project. Anyone can download the source code to run their own dedicated servers, and they can also modify, customize, and extend the functionality of the system. The project is distributed under the MIT license\footnote{\scriptsize{http://opensource.org/licenses/MIT}}, which is permissive with regard to both modifications and commercialization. MIT has made available a set of documents on how to extend the system, and development team members often hold remote video sessions in which they present on technical topics and in which community members are invited to show off their work. Members of the open source developer community have made many contributions to the project, ranging from small bug fixes, to full components such as the \emph{Slider}, a component that wraps the Seekbar\footnote{\scriptsize{http://developer.android.com/reference/android/widget/SeekBar.html}} widget available in the Android SDK.
\paragraph{}
The App Inventor development environment is supported for Mac OS X, GNU/Linux, and Windows operating systems, and the resulting apps can be installed on any Android phone running Android 1.5 (Cupcake) or newer.

\section{Research Projects}

\subsection{Computational Thinking Through Mobile Computing}

\paragraph{}
\emph{Computational Thinking Through Mobile Computing}\footnote{\scriptsize{http://explore.appinventor.mit.edu/mobileCT}} is a project funded by the National Science Foundation\footnote{\scriptsize{This material is based upon work supported by the National Science Foundation under Grant Numbers 1225680, 1225719, 1225745, 1225976, and 1226216}} for introducing students to computational thinking (CT) through creating mobile apps. Our five-university partnership (MIT, University of Massachusetts Lowell, University of San Francisco, Trinity College, and Wellesley College) is developing new materials and techniques for teaching CT through mobile app development. The research component of the project will illuminate the effectiveness of these new materials and techniques, and will culminate with the dissemination of project artifacts and findings.

\paragraph{}
New materials being developed include such things as a web-based collection of video lectures, screencasts, tutorials, programming exercises, quizzes, live-coding programming challenges, and concept maps. Modular materials are being created for use in a variety of undergraduate courses with plans to pilot materials in both computer science courses and interdisciplinary courses.
 
\paragraph{}
New techniques for assessing CT comprise a set of pre-surveys and post-surveys, grading rubrics for student projects, and automatic program analysis tools to assess student CT learning and validate these techniques. 

\subsection{Analysis of Program Structure in App Inventor Projects}

\paragraph{}
Although over a million App Inventor users have created about 2.5 million apps, little is known about the nature of these apps and the problems encountered by the people creating them. Research is underway to analyze the structure of App Inventor programs in order to gain insight into the effectiveness of this visual language for creating apps and for learning programming concepts \citep{turbak:analysis}.

\paragraph{}
As part of this research project, the team will develop notions of program sophistication in order (1) to determine the extent to which App Inventor users are learning programming concepts and (2) to provide feedback to users about their programs. These notions could be used as the basis for an automatic tutor for improving coding style. 

\paragraph{}
In addition to focusing on the structure of App Inventor programs, it is also possible to track program execution on devices, including runtime errors. This information will help us better understand the proficiencies and confusions of users and will also allow us to provide debugging support when an error occurs. 

\paragraph{}
A longer range goal is to instrument App Inventor to record fine-grained steps of program construction. This will give a more detailed narrative of how users write their programs. We expect that the high level nature of blocks programming edits will be more manageable and informative than the low-level nature of keystroke edits used in learning analytics systems for programming.

\subsection{Sensor Integration and Harnessing Personal Data}

Another research project underway is focused on how data from integrated sensors can inform teaching practice \citep{shih:sensors}. Researchers created a framework to collect, monitor and analyze student behavioral data and increase social interactions for educational uses. 

The framework has three essential components. The first component is App Inventor, which allows novices to create mobile apps that collect personal and environmental data. The second component is the Reactive Data Store, which was developed as an intuitive mechanism that connects multiple data sources and automates the tasks for analyzing personal data. The third component is a push notification mechanism that delivers personalized or group messages as the results of the real-time analysis and triggers more actions accordingly. Together, these components will transform people\textquoteright s campus experiences by unleashing the value of smartphone data. 

As part of the preliminary study, researchers introduced the framework to a group of higher-education teachers through a focus group discussion to identify the plausible use cases in terms of performance assessments, pedagogy design and collaborative learning. Results are forthcoming after more investigation.

\section{Evolution and Future of App Inventor}
\paragraph{}
Since moving to MIT, the App Inventor project has evolved to include an extensive collection of online support materials, an active online user community, an extended development team, and an extensive network of over one million users. The most recent major achievement for the project was moving the  Blocks Editor into the browser. The editor was originally written in Java, and run as a \emph{Java Web Start} application launched from the web-based Designer window. Within eighteen months of taking over the project, the MIT team rewrote the Blocks Editor using the Blockly\footnote{\scriptsize{https://code.google.com/p/blockly/}} library. The service will be relaunched as \textquotedblleft App Inventor 2\textquotedblright , making it entirely browser-based and eliminating many technical issues with running the software.

\paragraph{}
It is surprising how strongly computer programming is associated with plain black and white text on a screen. App Inventor and other visual programming languages like it offer a solid, friendly, and rewarding entree into manipulating technological machines like mobile devices, robots, and PCs. Team members, both at MIT and at partner universities, are currently pursuing several strands of research based on App Inventor. These projects are varied and cover topics such as data mining, sensor integration and manipulation, embedded systems, curriculum development, and online learning. One of the primary goals for the team in the coming year is to understand better how people learn with App Inventor.




\bibliographystyle{abbrvnat}


\end{document}